\documentstyle[12pt]{article}

\begin{document}

\title{Critical Crashes?} 
\author{ Kirill Ilinski \thanks{E-mail:  kni@th.ph.bham.ac.uk} \\
[0.3cm] 
{\small\it IPhys Group, CAPE,
14-th line of Vasilievskii's Island, 29}\\
{\small\it St-Petersburg, 199178, Russian Federation}
\\ [0.1cm]
{\small and} \\ [0.1cm]
{\small\it School of Physics and Space Research,
University of Birmingham,} \\
{\small\it Edgbaston B15 2TT, Birmingham, United Kingdom}
}

\vspace{3cm}

\date{ }

\vspace{1cm}

\maketitle

\begin{abstract} 
\noindent
In this short note we discuss recent attempts to
describe pre-crash market dynamics with analogies from 
theory of critical phenomena. 
\end{abstract}

Stochastic dynamics is a commonly used approach to describe the time
evolution of financial markets. The large number of agents, their complex
mutual influence and obscure decision making - all of these leave the
probabilistic picture as the only viable way to describe and, as a test,
to predict financial markets. In the case of a stable market such
a
probabilistic description is on solid ground. Indeed, for the stable
market the difference between the numbers of ``buy" orders and ``sell" orders is
small and fluctuates around zero which results in corresponding
fluctuations of the price of the traded asset. It means that peculiarities
of individual decisions are mutually offset, the characteristic fluctuation
time is less than the characteristic time of changes of external
(fundamental) conditions and number of observations allows
meaningful use of statistical procedures such as averaging.
However the situation changes dramatically when the overwhelming number of
market participants share the same view on the future and make the same
orders (for example, order to sell) thus creating a market instability.
The participants start to behave ``coherently" and to create a realistic
picture of such market one needs to model the decision making leading to
such behaviour. The modelling of this decision making is the principle
problem since the corresponding price movements are simple consequence
of the demand-supply mechanism of price fixing. The ``coherence" of such
events and their extremely low frequency make a statistical description
dubious and the whole problem even more challenging. Taking into account
the
immense importance of crashes as well as ``bubbles" for financial systems
and the world economy as a whole it is not surprising that the problem of
description and prediction of the events attract a lot of efforts and
attention.  In this note we concentrate on the recently suggested analogy
between critical phenomena and the financial crashes.

The story started with empirical observations of so-called
log-periodical oscillations prior the crashes which have been reported
in physical literature by several authors
\cite{S1,Feigenbaum,Vandewalle,DS}. The empirics was questioned later in
Ref.\cite{Bouchaud} where the stochastic nature of crashes was advocated
(see also \cite{Bouchaud1} on stochastic model of crashes).
We carried out an extensive search for the patterns for Russian
securities market and some other emerging markets and did not find
convincing manifestations. Though this itself does not undermine the
empirics and the developed theory suggested for well developed liquid
markets, it cast a shadow on implied universality of the effect.
Even with the most successful examples demonstrated in Ref.\cite{S0} (see
figs.1-3 therein) the log-periodic oscillations themselves generate
such timing of the ``sell" signals that would result in losses because of
time lag (generally a few months). In fact, as it can be seen from the
figures, the disciplined following of this strategy would result in up to
50\%
losses. This prompted a statistical (probabilistic) modification of
the ``sell" signals \cite{S0} which however significantly reduced
the
practical relevance of the method because of the lack of strict rules which
are so important in technical trading. One can argue that similar or
even better accuracy can be achieved by an investor with a common sense
who observes the growing market ``bubble". It does not mean that the
oscillation pattern is not worth to look at but rather that the real life
of financial market once again is more complicated and we still have a
long way to go to understand it.

It would be unfair to say that people never observed or used such
patterns. The theory of cycles constitutes a valuable part of the
Technical Analysis which aims to predict future market movements based on
the price history. All reported technical rules are never 100\%
accurate and hence they are probabilistic in their nature. The belief is
that they are right more often than wrong. Before using trading
strategies they are to be thoroughly back tested which does not guarantee
however the future performance. The environment changes all the time and
this can kill the profitability of a particular strategy. Nevertheless
practitioners tend to use technical predictions much more often than
the
recipes of canonic financial economics. For many years a huge number of
people have been involved in pattern studies and many text books are
available to cover the subject. No surprise then that it is possible to
find log-periodic oscillations there! The chartists use another language
and call it Log spirals to encapsulate both cyclic nature and the scaling
property (see \cite{SC} for an introduction and further references on the
subject) but the phenomena under study is essentially the same. There
exists a speculation that the Log spirals can be applied not only to
very large market movements such as crashes but also to less significant
market movements. A recent paper \cite{W} reported in physical literature a
similar observation.

Now let us turn to the theory of the 
log-periodic oscillations suggested in Refs.\cite{S2,S0}
(for recent developments see~\cite{S3}). 
It was postulated that the microscopic origin of the
phenomena is in mutual imitations by traders. This model and
the market fitting show that the average number of interactions $\delta -1$ 
felt by
a typical trader is order of $2\sim 3$ but the traders are so
mixed up that "long" connections can appear. This indeed resembles the
well known mechanism of Long Range Order formation in spin systems
which is the main attraction as well as the main danger for the
application of the model to financial markets. The model is based on the
assumption that the trader-trader imitation is the principle factor
which governs the dynamics. Though the imitation is definitely an important
factor it is hard to believe that it is the principle one.  Indeed,
traders are clustered and it is not very realistic to imagine that the
clusters are so intertwined. The market participants have very different
time horizons (like minutes for speculators and years for fund managers)
and the assumption that they watch each other for years prior the crashes
does not seem to be quite motivated. At the same time there are several
sources of information which influence most of the traders.  Reuters news
for institutional traders, the Financial Times or BullMarket reports for
small investors play major role to form their expectations.  The
``coherence" might be a result of a common factor action rather than  a
result of a direct internal interaction such as the imitation of 2-3
fellow traders. It does not mean however that the news impact is
something external since the news can reflect the market anticipations
and, hence, represents some indirect impact. In any case it is more
realistic to expect that common information resources will be more
relevant to define investment decisions and create the ``coherence" than
the obscure network of fellow traders.

The microscopic theory gave the probability
per unit time that the crash will happen in the next moment if it
has not happened yet (hazard rate) $h(t)$ as defined by the "mean
field" equation:
$$
\frac{d h}{d t}= const \cdot h^{\delta}  \ .
$$
Furthermore, it was assumed that the corresponding price $p$
follows the stochastic process
$$
dp = \mu(t) p(t) - k [p(t) - p_1] dj
$$
where $k$, $p_1$ are some constants and
$j$ is a jump process whose value is zero prior the crash and one
afterwards and which is characterized by the hazard rate.
To obtain an equation for the return $\mu(t)$ authors postulate the
following ``fair" game (or martingale) hypothesis:
$$
E_t (p(t^{\prime})) = p(t) \ , \quad t< t^{\prime} \ .
$$
Here $E_t$ denotes an expectation conditional on information revealed up
to time $t$. This is the last equation which we discuss now and show
that it cannot be used in the situation of a financial ``bubble" prior to the
crash.

The martingale property for the price is associated with the market efficiency.
It would be wrong to say however they are equivalent. 
Let us remind that  the Efficient Market Hypothesis is defined 
as a superposition of the Rational Expectation Hypothesis and Orthogonality
property~\cite{Cuthbertson}. The {\it Rational Expectation Hypothesis} states 
that:
\begin{enumerate}
\item Agents are rational, i.e. use any possibility to get more than less if the 
possibility occurs.
\item There exists a perfect pricing model and all market 
participants know this model.
\item Agents have all relevant information to incorporate into the model.
\end{enumerate}
Using the model and the information the rational agents form an expectation 
value of the future return $E_t R_{t+1}$. This expectation value can differ 
from the actual value of the return $R_{t+1}$ on a estimation error
$\epsilon _t = R_{t+1} - E_t R_{t+1}$. The {\it Orthogonality property} 
implies that:
\begin{enumerate}
\item $\epsilon _{t+1}$ is a random variable which appears due to coming of
new information.
\item $\epsilon _{t+1}$ is independent on full information set $\Omega _t$ 
at time $t$ and 
$$
E_t (\epsilon _{t+1}|\Omega _t) =0 \ .
$$
\end{enumerate}
The martingale property for the price then appears from the Orthogonality
property under the additional assumption that the perfect pricing model gives 
zero return as a best prediction:
\begin{equation}
0=E_t (\epsilon _{t+1}|\Omega _t) = E_t (R_{t+1}|\Omega _t) =
E_t (p_{t+1} - p(t)|\Omega _t)   \Longleftrightarrow 
E_t (p_{t+1}|\Omega _t)= p(t) \ .
\label{e}
\end{equation}
This means that the model of crashes we discussed above assumes a zero
return as a best prediction for the market. No need to say that this is 
not what one expects from a perfect model of market ``bubble"! 
Buying shares traders expect the price to rise and it is reflected 
(or caused) by their prediction model. They support the "bubble" and the 
``bubble" support them! This is not a coin tossing as Eqn(\ref{e}) suggests.
In this situation  the whole market on average makes money rather than 
redistributes them as it is in fair game. 
These expectations have to be incorporated in the prediction model and indeed
there exist several models of so-called
rational ``bubbles" which are models to describe ``bubbles" under of the 
Efficient Market Hypothesis. We address to the 
textbook~\cite{Cuthbertson} (Chapter 7)
for a  review or to the original papers~\cite{bub1,bub2} for further reading.

It is possible to look at the problem with Eqn(\ref{e}) from other side. 
The equation postulates that the average price tomorrow is equal to the price 
today, i.e. the average rate of return is zero. It is well-known that, 
in general, investors are risk averse and require some additional risk premium 
to enter a bet. Otherwise they simply stay out and do not take  the risk.
The unrealistic feature of Eqn(\ref{e}) is that it takes the risk premium
equal to zero. One can improve the situation substituting Eqn(\ref{e}) by the
following equation:
\begin{equation} 
\nu (t) E_t (p_{t+1}|\Omega _t)= p(t) \ .
\label{ee}
\end{equation}
where $\nu (t)$ is an appropriate discount factor which reflects the risk premium 
required by market participants. A common and often used approximation,
$\nu (t) = $const, does not work well for  the problem of ``bubbles" and 
crashes (see also discussion in \cite{Soros}). This is due to the fact that the
risk premium is defined by risk perceptions which are constantly changing in the 
course of the ``bubble" according to a prediction model adopted by investors.
Furthermore, to model the discount factor dynamics one needs once again a 
prediction model. At this point we return to the previous consideration.

It is an open question how to construct a perfect model or how to model 
traders decisions. Computer agents simulations give one of possible approaches.
This is a new developing area and the last decade many research papers addressed 
the question~\cite{Agents}. It is difficult to say how accurate is the modelling
and, hence, how reliable are the models. They mostly consider a market as a mixture
of technical and fundamental traders and prescribe the chartists a set of 
technical trading rules. It is important to emphasize that there is no 
contradiction in using the technical analysis as a proxy of the perfect model 
under EMH and the Market Efficiency itself (see~\cite{II21} for more discussion). 
All relevant effects such as risk aversion, 
existence of common news source, the mentioned earlier common sense and 
imitations, even the log-periodic oscillations themselves can be included 
in the model.

In conclusion, the discussed above log-periodic oscillations or Log
spirals can be considered as one of technical instruments to predict
market movements. As any market phenomena they are worth to study both
empirically and theoretically. To make the method more practically
applicable it is important to combine it with other technical methods to
validate the trading signals which is a common practice in Technical
trading since the performance of the method alone does not produce
convincing results. Finding a realistic theoretical model (not
necessarily triggered by physical analogies) is definitely a difficult and
challenging problem where the only the first steps have been made.

\section*{Acknowledgments}
I am grateful to Didier Sornette for his valuable comments on
earlier version of the paper. I also really appreciate a number of enjoyable 
and revealing discussions of the subject which I had during last year.


\begin{thebibliography}{99}

\bibitem{S1} D. Sornette, A. Johansen and J.P. Bouchaud, {\it
J.Phys.I (France)}, {\bf 6} 167-175 (1996);

\bibitem{Feigenbaum} J.A. Feigenbaum and P.G.O. Freund {\it Int.J.Mod.Phys.}
{\bf B 10}, 3737-3745 (1996);

\bibitem{Vandewalle} N. Vandewalle, M. Ausloos, Ph. Boveroux, A. Minguet,
{\it Eur.Phys.J} {\bf B 4}, 139-141 (1998); {\it Physica} {\bf A 255}, 
201-210 (1998);

\bibitem{DS} D. Sornette and A. Johansen, 
{\it Physica } {\bf A 245}, 411-422 (1997); 

\bibitem{Bouchaud} L. Laloux, M. Potters, R. Cont, J.-P. Aguilar and 
J.P. Bouchaud, {\it Europhys. Lett.} {\bf 45}, 1-5 (1999);

\bibitem{Bouchaud1}  J.P. Bouchaud and R. Cont, {\it Eur.Phys.J.} 
{\bf B 6}, 543-550 (1998);

\bibitem{S0} A. Johansen and D. Sornette, {\it Risk} {\bf 12}, 91-94 (1999); 

\bibitem{SC} W.T. Erman: "Log Spirals in The Stock Market", {\it
Technical Analysis of Stock \& Commodities}, {\bf 17}, 16-34 (1999);

\bibitem{W} S. Drozdz, F. Ruf, J. Speth and M. Wojcik:
"Imprints of Log-periodic self-similarity in the stock market",
preprint cond-mat/9901025, available at 
http://xxx.lanl.gov/abs/cond-mat/9901025;

\bibitem{S2} A. Johansen, O. Ledoit and D. Sornette: "Crashes as critical 
points", preprint cond-mat/9810071, available at 
http://xxx.lanl.gov/abs/cond-mat/9810071;

\bibitem{S3}
A. Johansen and D. Sornette: 
``Modeling the Stock Market prior to large crashes", 
to appear in {\it European Physical Journal} {\bf B};
``Financial ``anti-bubbles'': log-periodicity in Gold and Nikkei collapses", 
to appear {\it Int. J. Mod. Phys.} {\bf C};

\bibitem{Cuthbertson} K.  Cuthbertson, {\it Quantitative financial
economics}, John Wiley $\&$ Sons, 1996;

\bibitem{bub1} 
O.J.Blanchard: "Speculative Bubbles, Crashes and Rational expectations", 
{American Economic Review}, {\bf 81}, 1189-1214 (1991);

\bibitem{bub2} 
K.A. Froot and M. Obstfeld: "Intrinsic Bubbles: The Case 
of Stock Prices", {\it American Economic Review}, {\bf 81},
1189-1214 (1991);

\bibitem{Soros} George Soros, {\it Alchemy of Finance}, 
John Wiley $\&$ Sons, 1987;

\bibitem{Agents}  G.W. Kim and H.M. Markowitz, 
{\it J. Portfolio Management}, {\bf 16}, 45 (1989);
M. Levy, H. Levy and S. Solomon, {\it Economics Letters},
{\bf 45} 103-111 (1994); {\it J.Phys.I (France)} {\bf 5}, 1087-1107 (1995);
R.G. Palmer, W.B. Arthur, J.H. Holland, B.LeBaron and P.Taylor, {\it Physica}
{\bf D 75}, 264-274 (1994); P. Bak, M. Paczuski and M.Shubik, {\it Physica}
{A 246}, 430-453 (1997); G. Caldarelli M. Marsili and Y.-C. Zhang, 
{\it Europhys. Lett.} {\bf 40} 479-484 (1997); T. Lux and M. Marchesi,
{\it Nature} {\bf 397}, 498-500 (1999); 

\bibitem{II21} A. Ilinskaia and K. Ilinski:
"How to reconcile Market Efficiency and Technical Analysis", 
IPhys Group working paper, preprint cond-mat/9902044,
available at 
http://xxx.lanl.gov/abs/cond-mat/9902044;

\end{thebibliography}
\end{document}